\documentclass[cameraready]{Interspeech}

\usepackage{algorithm}
\usepackage{algorithmic}
\usepackage{amsmath,amsfonts}
\usepackage{placeins}

\title{Which Speech Representation Better Matches Text-Native Reasoning? A Study of Speech-Text Alignment on Frame Rate and Representation}

\author[affiliation={1}]{Zhen}{Ye}
\author[affiliation={2}]{Xu}{Tan}
\author[affiliation={1}]{Yiming}{Li}
\author[affiliation={7}]{Guangyan}{Zhang}
\author[affiliation={1}]{Chimin}{Chan}
\author[affiliation={3}]{Haohe}{Liu}
\author[affiliation={4}]{Zhengxi}{Liu}
\author[affiliation={5}]{Hongzhan}{Lin}
\author[affiliation={4}]{Zheqi}{Dai}
\author[affiliation={6}]{Xinshen}{Zhang}
\author[affiliation={4}]{Peiwen}{Sun}
\author[affiliation={4}]{Qiuqiang}{Kong}
\author[affiliation={1},correspondingauthor]{Wei}{Xue}

\address{
    $^1$ Hong Kong University of Science and Technology, Hong Kong SAR \\
    $^2$ Tencent, China \\
    $^3$ University of Surrey, United Kingdom \\
    $^4$ Chinese University of Hong Kong, Hong Kong SAR \\
    $^5$ Hong Kong Baptist University, Hong Kong SAR \\
    $^6$ Hong Kong Polytechnic University, Hong Kong SAR \\
    $^7$ Independent Researcher
}

\email{zhenye213@gmail.com, weixue@ust.hk}


\begin{document}

\maketitle

\begin{abstract}
Spoken dialogue models typically start from text LLM backbones, yet reasoning often degrades when conditioning on speech instead of text. We attribute part of this modality gap to a temporal-granularity mismatch: speech tokens are temporally redundant and far longer than text under matched semantics, diluting per-token semantic density and weakening text-native reasoning dynamics. We study speech token design as a representation selection problem and sweep frame rates under a frozen LLM backbone with a fixed information rate. To make low frame rates feasible, we introduce factorized FSQ and a lightweight non-autoregressive audio LM head, scaling capacity to nearly 300\,bits/frame without sacrificing efficient prediction. With the bottleneck removed, we sweep frame rates (50$\rightarrow$2.08\,Hz) and alignment depth, and observe a consistent best regime for speech QA at 4.17\,Hz with intermediate-layer representation alignment.
\end{abstract}

\section{Introduction}

End-to-end spoken dialogue models built on text LLM backbones~\cite{defossez2024moshi,xu2025qwen2,ding2025kimi,huang2025step} have demonstrated impressive capabilities, yet a persistent \emph{modality gap} remains: reasoning quality degrades when the model conditions on speech tokens rather than text tokens. A common response is to narrow this gap by end-to-end post-training the backbone on speech or speech--text mixtures, allowing the model to adapt its internal dynamics to speech inputs. However, this approach is costly and, more importantly, entangles two factors: improved speech representations versus changes in the LLM itself. As LLM scale grows, such full-model adaptation makes it increasingly difficult to diagnose \emph{why} speech conditioning degrades reasoning and which representation choices actually matter. In this work, we instead freeze the text LLM and treat speech token design as a representation selection problem under a controlled setup.

In the frozen-LLM setting, a key and under-explored contributor is a \emph{temporal-granularity mismatch}. Typical speech tokenizers operate at 12.5--50\,Hz, producing sequences an order of magnitude longer than text for the same utterance. As shown in Fig.~\ref{fig:ratio_dist}, the average text token rate on LibriSpeech is only ${\sim}$3.32\,Hz---meaning a 50\,Hz speech tokenizer generates roughly 15$\times$ more tokens than necessary to convey the same semantic content. This temporal redundancy dilutes per-token semantic density, burdens the self-attention mechanism with many low-information positions, and perturbs the text-native reasoning dynamics the LLM was pretrained with.
Since text LLMs are pretrained to model a single discrete stream with a single next-token prediction head, a natural starting point is a single-codebook speech stream predicted by a single LM head under the same objective. We take this text-native design as the default baseline, and ask what minimal modifications are needed to reduce frame rate toward the text token rate without losing information.
Frame rate is therefore a first-class lever: reducing it increases semantic density and shortens the sequence, but pushing it too low forces each token to carry substantially more information and exposes severe bottlenecks under standard single-codebook, single-head designs. We address this low-rate bottleneck with factorized finite scalar quantization (FSQ) and a lightweight non-autoregressive audio LM head, scaling capacity to nearly 300\,bits/frame while keeping prediction efficient.

In this paper, we investigate what speech tokenization best aligns with text tokens so as to inherit the reasoning capabilities of pretrained text LLMs. We do so under a controlled setup where the LLM backbone is kept fixed and the information rate (bits/second) is held constant, so performance differences can be attributed to the speech representation itself.
We study this from two perspectives. First, from the perspective of \emph{frame rate}, we compare against the average text token rate of 3.32\,Hz (Fig.~\ref{fig:ratio_dist}) and systematically test speech frame rates from 50\,Hz down to 2.08\,Hz under a fixed information rate---covering regimes where speech sequences are longer than, comparable to, or shorter than text. Systematically exploring such low-rate tokenizations is challenging because severe information bottlenecks naturally emerge under standard single-codebook, single-head designs; to enable this sweep, we propose a new framework that mitigates information loss at very low frame rates and enables efficient parallel prediction.
Second, from the perspective of \emph{representation alignment}, we introduce a contrastive objective that explicitly encourages speech and text embeddings to lie in a shared latent space at intermediate LLM layers, thereby closing the cross-modal semantic gap.

Our contributions are threefold:
\begin{enumerate}
    \item \textbf{Length alignment and new architecture.} We systematically explore speech token frame rates from 50\,Hz to 2.08\,Hz under a fixed information rate. To overcome the information bottleneck at low frame rates, we propose a scalable architecture with factorized FSQ codebooks and a lightweight NAR audio LM head.
    \item \textbf{Representation alignment.} We introduce intermediate-layer contrastive alignment via InfoNCE, showing that mid-layer alignment is significantly more effective than embedding-level or late-layer alignment.
    \item \textbf{Empirical insights for speech token design.} Our experiments show that a frozen text LLM with only ${\sim}$100M trainable parameters and 2.5k hours of data achieves competitive speech-to-speech QA. The findings provide practical guidance for future speech token design, revealing how frame rate and alignment depth jointly affect cross-modal transfer.
\end{enumerate}

\section{Related work}

\subsection{Spoken dialogue modeling}

Our work studies which speech representation best preserves text-native reasoning in LLM-based spoken dialogue. We briefly review the two dominant paradigms and the frozen-LLM line most relevant to our study.

\textbf{Cascaded systems} decompose the problem into ASR, text LLM, and TTS modules. This modularity fully preserves the text LLM's reasoning capability but incurs high latency from serial processing, suffers from error propagation across stages, and discards paralinguistic cues (e.g., emotion, prosody) during the intermediate text stage.

\textbf{End-to-end systems} align speech representations directly with LLM hidden spaces. SpeechGPT~\cite{zhang2023speechgpt} progressively adapts LLMs to quantized self-supervised speech tokens. LLaMA-Omni~\cite{fang2024llama} employs a streaming decoder with CTC-based simultaneous generation. Interleaved methods like Spirit-LM~\cite{nguyen2025spirit} and GLM-4-Voice~\cite{zeng2024glm} alternate between text and speech tokens within a single sequence. Parallel paradigms such as Mini-Omni~\cite{xie2024miniomni} and SLAM-Omni~\cite{chen2024slamomni} model text and speech through dual-head architectures. These systems universally operate at high speech frame rates (12.5--50\,Hz), producing sequences far longer than the corresponding text---a temporal-granularity mismatch whose impact on reasoning transfer has not been systematically studied.

\textbf{Frozen-LLM approaches.} Freezing the text LLM isolates the speech representation as the sole variable, making it ideal for our controlled study. BLSP~\cite{wang2024blsp} aligns speech encoders to frozen LLMs via behavior alignment of continuation writing. AudioChatLLaMA~\cite{fathullah2024audiochatllama} extends this to multi-modal dialogue. However, both focus on speech understanding only. DIVA~\cite{held2024diva} applies contrastive alignment at the input embedding layer. None of these works investigate frame rate as a design variable, nor do they align at intermediate hidden layers or support pure speech-token generation.

\subsection{Speech tokenizer}

Since our study treats speech tokenization as the key representation choice, we review tokenizer designs with attention to their frame rate and quantization structure---the two factors our controlled sweep varies.

Acoustic tokenizers derived from neural codecs such as EnCodec~\cite{defossez2022encodec} and DAC~\cite{kumar2023dac} reconstruct waveforms with high fidelity using multi-layer RVQ~\cite{lee2022rvq} at high frame rates (e.g., 75\,Hz). Semantic tokenizers prioritize linguistic content: SpeechGPT~\cite{zhang2023speechgpt} uses HuBERT~\cite{hsu2021hubert} tokens; SLAM-Omni~\cite{chen2024slamomni} and GLM-4-Voice~\cite{zeng2024glm} quantize Whisper features; Moshi~\cite{defossez2024moshi} combines RVQ with semantic distillation; SpeechTokenizer~\cite{zhang2024speechtokenizer} disentangles semantic and acoustic information across RVQ layers. All these designs operate at 25--50\,Hz, producing speech sequences substantially longer than text---yet none study how reducing this rate toward the text token rate affects LLM reasoning transfer.

Regarding codebook design, the concept of grouped or factorized quantization originates from VQ-VAE~\cite{vandenoord2017vqvae} and is extended in NaturalSpeech3's FACodec~\cite{ju2024naturalspeech3}. We apply factorized FSQ prediction at extreme low frame rates (down to 2.08\,Hz) specifically to study frame-rate alignment with frozen LLMs. For cross-modal alignment, DM-Codec~\cite{lyu2024dmcodec} aligns speech and text at selected LLM layers using cosine similarity. We instead use InfoNCE, which provides batch-level contrastive learning with harder negatives. Unlike prior tokenizer studies that focus on reconstruction quality or ASR performance in isolation, we study the joint interaction of frame rate, codebook capacity, and alignment depth across the full understanding-to-generation pipeline.

\section{Length alignment}

\subsection{Preliminary}

\begin{figure}[t]
    \centering
    \includegraphics[width=1.0\linewidth]{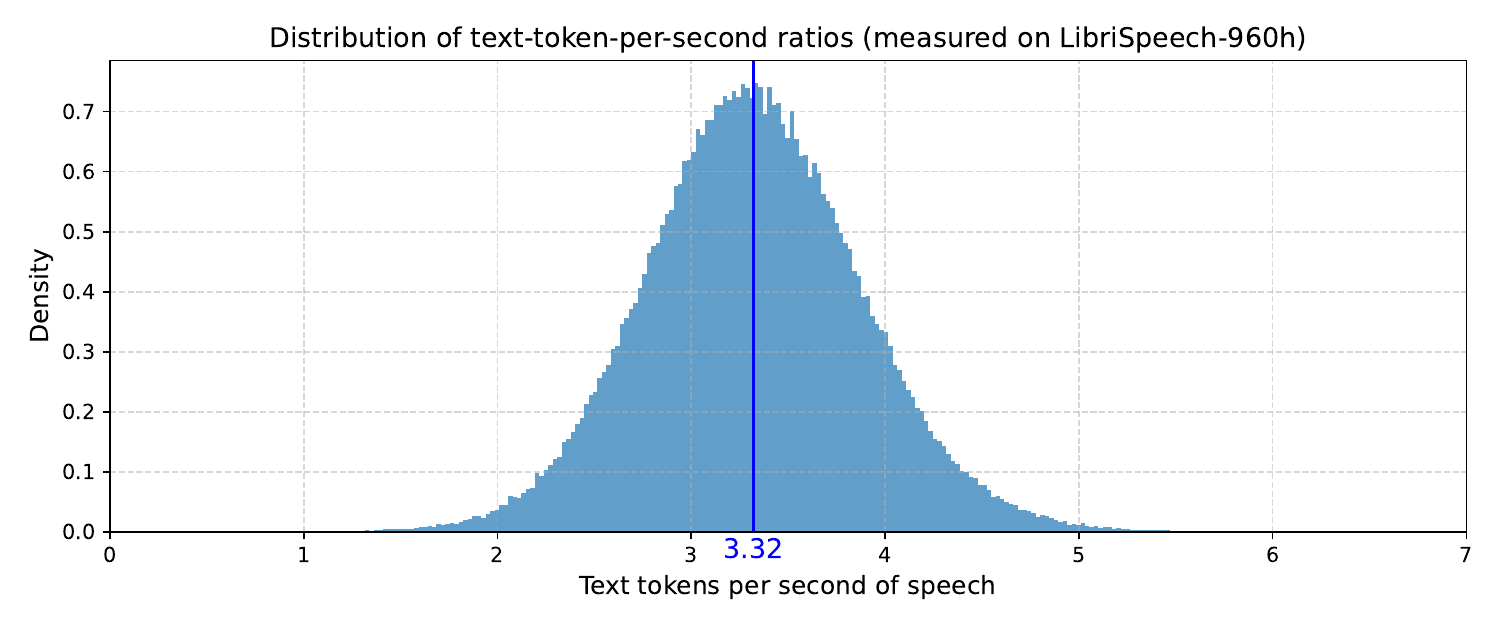}
    \caption{Distribution of text-token-per-second ratios on LibriSpeech-960h, tokenized with Qwen3 tokenizer. Blue line: mean (3.32\,Hz).}
    \label{fig:ratio_dist}
\end{figure}

We use discrete speech tokens rather than continuous representations, as discrete tokens allow speech to be treated identically to text under the same autoregressive next-token prediction paradigm, enabling direct reuse of the text LLM's generation machinery without modifying the architecture.

We begin by examining the length ratio between speech and text tokens. Using LibriSpeech-960h~\cite{panayotov2015librispeech} (${\sim}$256k utterance-transcript pairs), we compute the ratio of text token length to speech duration. Transcriptions with preserved capitalization and punctuation are taken from LibriSpeech-PC~\cite{meister2023librispeech}; text is tokenized with the Qwen3 tokenizer~\cite{yang2025qwen3}. As shown in Fig.~\ref{fig:ratio_dist}, the average ratio is approximately 3.32 tokens/second.

\begin{figure}[t]
    \centering
    \includegraphics[width=1.0\linewidth]{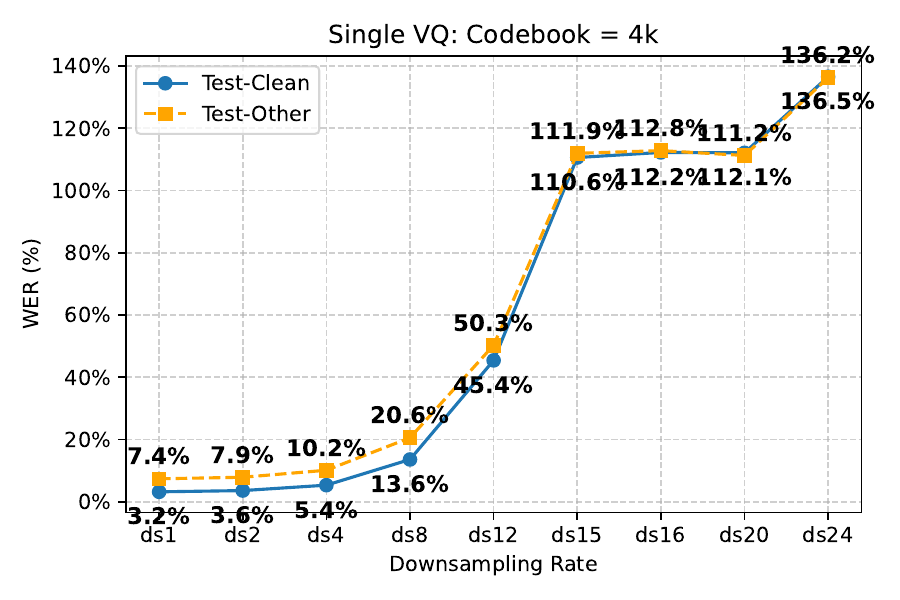}
    \caption{ASR WER vs.\ downsampling rate with a 4k codebook ($2^{12}$) on LibriSpeech test-clean/test-other.}
    \label{fig:singlevqwer_4k}
\end{figure}

\begin{figure}[t]
    \centering
    \includegraphics[width=1.0\linewidth]{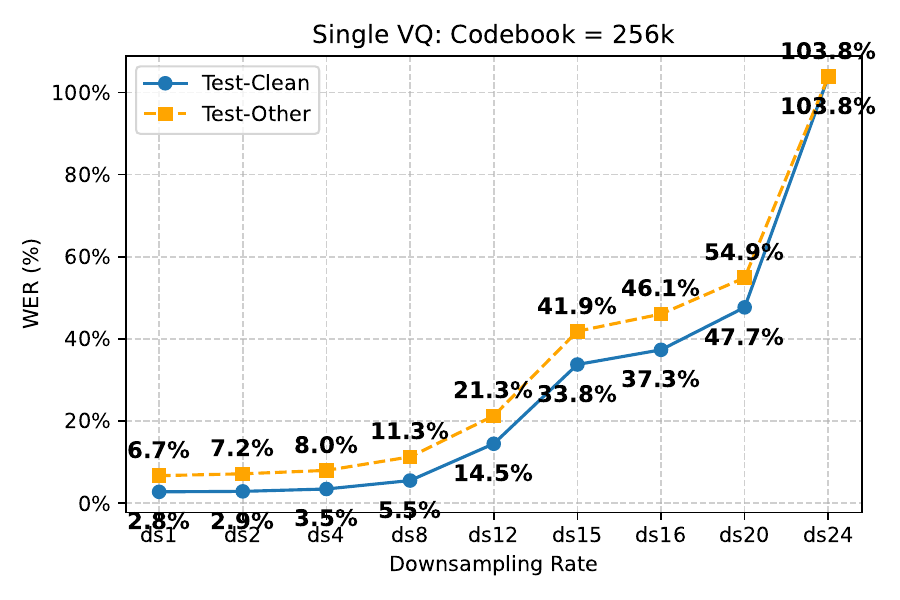}
    \caption{ASR WER vs.\ downsampling rate with a 256k codebook ($2^{18}$) on LibriSpeech test-clean/test-other.}
    \label{fig:singlevqwer_256k}
\end{figure}

This naturally raises a question: does matching the speech token frame rate to the text token rate actually improve alignment? Unfortunately, prior architectures~\cite{zeng2024glm,zhang2023speechgpt,nguyen2025spirit,hassid2023textually,lakhotia2021generative} universally adopt single-VQ + single-LM-head designs. When the frame rate is low, each speech token must encode a large amount of acoustic and linguistic information into a single codebook entry, creating an \emph{information bottleneck}. We verify this with a controlled ASR experiment: Whisper-Large-v3 features are downsampled at various rates, quantized with FSQ, projected into the frozen LLM embedding space via a learned linear mapping, and decoded autoregressively as text. As shown in Fig.~\ref{fig:singlevqwer_4k}, with a 4k codebook ($2^{12}$), ASR performance degrades sharply when the downsampling rate exceeds 4$\times$ (i.e., below 12.5\,Hz); even scaling the codebook to 262k ($2^{18}$) fails to maintain acceptable WER below a downsampling rate of 8$\times$ (6.25\,Hz), as shown in Fig.~\ref{fig:singlevqwer_256k}. This demonstrates that simply reducing the frame rate under standard architectures is insufficient---a fundamentally different approach to codebook design is needed.

\subsection{Proposed solution}

To overcome the information bottleneck, the most straightforward idea is to increase the codebook size. For speech \emph{input}, this is unproblematic: we only quantize features and project them into the text embedding space. However, for speech \emph{prediction}, increasing vocabulary size directly increases the LM head parameters and softmax computation, which becomes prohibitive at the codebook sizes required for low frame rates.

We instead exploit the factorized structure of FSQ~\cite{mentzer2023fsq}. In FSQ, each dimension of a $d$-dimensional feature vector is independently quantized to $L$ discrete levels, yielding an implicit codebook of size $L^d$ without requiring explicit codebook vectors or commitment losses. Rather than predicting from this exponentially large space in a single softmax, we partition the $d$ dimensions into $n$ groups, each covering $d/n$ dimensions with $L^{d/n}$ possible combinations. Each group is predicted in parallel, factorizing the intractable $L^d$-way classification into $n$ manageable $L^{d/n}$-way predictions. To condition each group's prediction on the same LLM output, we add a learnable \emph{slot embedding} (unique per group) to the LLM hidden state, allowing a \emph{shared} classification layer to serve all groups while maintaining group-specific context.

Formally, let $z_t\in\mathbb{R}^d$ be the downsampled speech feature at frame $t$, split into $n$ groups $z_t^{(g)}\in\mathbb{R}^{d_g}$ where $d = n d_g$. FSQ quantizes each scalar independently:
\begin{equation}
q_{t,g,k} = Q\left(z^{(g)}_{t,k}\right)\in\{0,1,\ldots,L{-}1\},\quad k=1,\ldots,d_g.
\end{equation}

Each group's discrete token is encoded as a mixed-radix (base-$L$) index over its $d_g$ scalar codes, which
factorizes an otherwise exponential $L^{d_g}$ vocabulary into $d_g$ small $L$-ary components for efficient
prediction and lookup.

\begin{equation}
y_{t,g} = \sum_{k=1}^{d_g} q_{t,g,k}\,L^{k-1}\in\{0,1,\ldots,K{-}1\},\qquad K=L^{d_g}.
\end{equation}
Thus each frame is represented by $y_t=(y_{t,1},\ldots,y_{t,n})$ with per-frame capacity $n\log_2 K$ bits (implicit codebook size $K^n=L^d$). At inference, the predicted group token $\widehat{y}_{t,g}$ is deterministically \emph{dequantized} back to the quantized continuous vector used by the speech decoder by first expanding it into scalar codes
\begin{equation}
\widehat{q}_{t,g,k} = \left\lfloor \widehat{y}_{t,g}/L^{k-1} \right\rfloor \bmod L,
\end{equation}
then mapping each scalar code to its quantized level value via the FSQ inverse mapping $\widehat{z}^{(g)}_{t,k}=Q^{-1}(\widehat{q}_{t,g,k})$, and finally concatenating all groups $\widehat{z}_t=\mathrm{concat}_g(\widehat{z}^{(g)}_t)$. This avoids learning an explicit embedding table over the exponentially large $L^d$ codebook.

To further increase the expressive capacity, we replace the standard linear classification head with a lightweight NAR transformer. Our NAR audio head consists of 2 transformer layers with hidden size matching the LLM backbone. The transformer processes all $n$ slot-augmented hidden vectors simultaneously through self-attention, allowing inter-group dependencies to be captured before the final shared classification layer.

Specifically, let $h_t\in\mathbb{R}^H$ denote the (frozen) LLM hidden state at the speech frame position. We construct group-specific queries by adding a slot embedding $s_g\in\mathbb{R}^H$:
\begin{equation}
u_{t,g} = h_t + s_g,\qquad g=1,\ldots,n.
\end{equation}
The NAR audio head $f_\theta$ attends over the $n$ group queries in parallel and outputs $v_{t,g}=f_\theta(u_{t,1:n})_g$. A shared classifier then predicts each group's token:
\begin{equation}
\begin{aligned}
\ell_{t,g} &= W v_{t,g} + b,\\
p(y_{t+1,g}=k \mid h_t) &= \mathrm{softmax}(\ell_{t,g})_k,
\end{aligned}
\end{equation}
where $k\in\{0,\ldots,K{-}1\}$.
We train the audio head with causal next-token prediction (teacher forcing): the hidden state at position $t$ predicts the group tokens at position $t{+}1$.
\begin{equation}
\mathcal{L}_{\text{audio}}=-\sum_{t=1}^{T-1}\sum_{g=1}^{n} \log p(y^\star_{t+1,g}\mid h_t).
\end{equation}

\begin{figure*}[t]
    \centering
    \includegraphics[width=\textwidth]{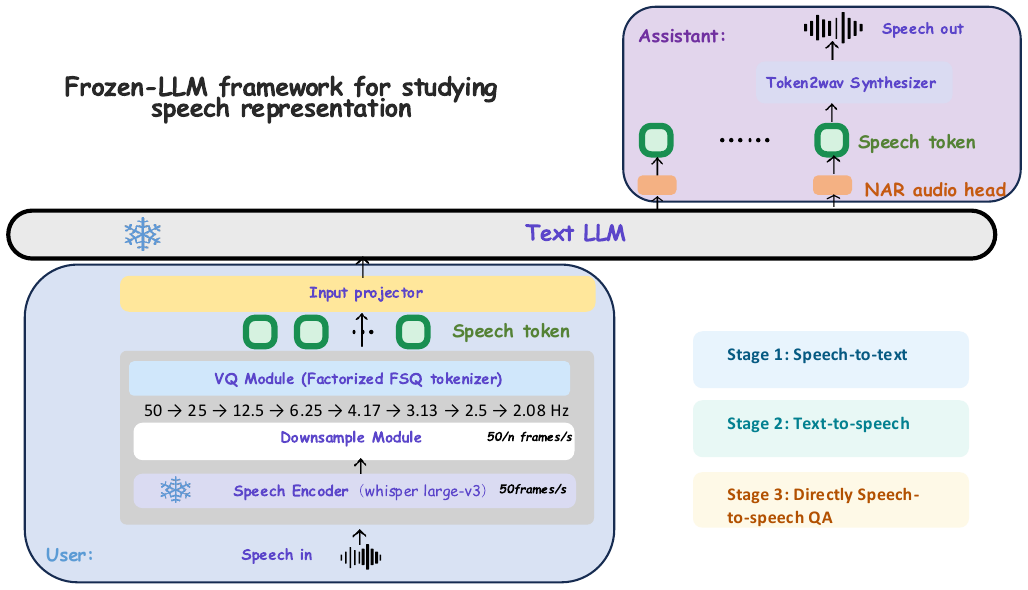}
    \caption{Architecture of our pure speech-token dialogue model. A frozen Whisper-Large-v3 encoder extracts 50\,Hz speech features, which are downsampled and quantized with (factorized) FSQ into grouped speech tokens, then projected into a frozen text LLM. For speech generation, a lightweight non-autoregressive audio head predicts the next-step speech tokens in parallel.}
    \label{fig:Arch}
\end{figure*}

Finally, we choose the number of groups $n$ as a function of the frame rate $r$ to maintain a constant information throughput
\begin{equation}
R = r\,n\,\log_2 K.
\end{equation}
In practice, we fix $K$ (per-group vocabulary) and vary $n$ accordingly. The overall architecture is illustrated in Fig.~\ref{fig:Arch}.

\section{Representation alignment}

After sequence length alignment, speech and text tokens may still reside in distinct latent subspaces of the LLM.
To address this, we introduce a representation alignment objective that explicitly encourages semantically corresponding speech and text embeddings to lie closer in the same hidden space.

A key design choice is that we operate on intermediate hidden states of the frozen LLM, rather than only on the embedding space. This layer-wise alignment allows semantic correspondence to be injected throughout the LLM hierarchy while keeping the backbone parameters fixed.
In practice, for each speech--text pair, we extract hidden states from both modalities at a selected LLM layer $\ell$, average over the temporal dimension, and obtain L2-normalized vectors $\widehat{h}_s, \widehat{h}_t \in \mathbb{R}^d$.
Since speech and text sequences generally differ in length (even after frame-rate alignment), we compute utterance-level representations via temporal average pooling over each modality's hidden states independently, which yields fixed-dimensional vectors regardless of input length. This assumes that the alignment between speech and text is roughly monotonic at the utterance level, which is a reasonable approximation for the paired data in our setup.
We then optimize a contrastive objective of the InfoNCE form:
\begin{equation}
\mathcal{L}_{\text{align}}
= - \sum_{i=1}^B
\log \frac{\exp\bigl( \langle \widehat{h}_{s,i}, \widehat{h}_{t,i} \rangle / \tau \bigr)}
{\sum_{j=1}^B \exp\bigl( \langle \widehat{h}_{s,i}, \widehat{h}_{t,j} \rangle / \tau \bigr)}
\end{equation}
where $B$ is the batch size, $\tau{=}0.07$ is a temperature hyperparameter, and $\langle \cdot, \cdot \rangle$ denotes the inner product (cosine similarity after normalization). The overall training loss combines task supervision and alignment:
\begin{equation}
\mathcal{L} = \mathcal{L}_{\text{CE}} + \lambda_{\text{align}} \, \mathcal{L}_{\text{align}}
\end{equation}
where $\mathcal{L}_{\text{CE}}$ is the standard next-token cross-entropy loss of the LLM, and $\lambda_{\text{align}}{=}0.1$ balances the contribution of the alignment term. Gradients from $\mathcal{L}_{\text{align}}$ flow only through the speech pathway (input projector and audio head); the frozen LLM backbone receives no gradient updates.

\section{Experiments}

We use Qwen3-4B~\cite{yang2025qwen3} as the frozen text LLM backbone and use Whisper-Large-v3 encoder as a frozen speech feature extractor, producing 50\,Hz representations. All backbone parameters are kept frozen; only the input projector and NAR audio LM head are trained. All configurations fix the information rate at 600\,bits/s. We test downsampling factors of 1$\times$ (50\,Hz), 2$\times$ (25\,Hz), 4$\times$ (12.5\,Hz), 8$\times$ (6.25\,Hz), 12$\times$ (4.17\,Hz), 16$\times$ (3.13\,Hz), 20$\times$ (2.5\,Hz), and 24$\times$ (2.08\,Hz).
This adds ${\sim}$100M trainable parameters for the 4B backbone and ${\sim}$150M for the 8B backbone (projector + audio head), while the text LLM remains entirely frozen.
We adopt three stages (S2T, T2S, and S2S QA) because they probe complementary aspects of alignment: speech-as-input understanding, speech-token generation, and the end-to-end combination required by spoken QA, enabling us to isolate the effect of frame rate and representation alignment under a frozen backbone.
\subsection{Training pipeline}

Our training follows a three-stage progressive pipeline.

\textbf{Stage 1: Speech-to-text (ASR).} The Whisper-Large-v3 encoder extracts 50\,Hz features from raw audio. These features are downsampled by a strided convolutional layer to the target frame rate, then quantized using FSQ. The quantized features are projected into the LLM embedding space via a learnable linear layer (the \emph{input projector}). The frozen LLM autoregressively generates text tokens from these projected speech embeddings. Only the downsampling layer, FSQ parameters, and input projector are trained.

\textbf{Stage 2: Text-to-speech (TTS).} The input projector is initialized from Stage~1, and the speech tokens established in Stage~1 are kept fixed as tokenizer. The frozen LLM receives text tokens as input and generates speech tokens via the NAR audio head. Only the input projector and NAR audio head are trained.

\textbf{Stage 3: Speech-to-speech QA.} Both the input projector and NAR audio head are initialized from their Stage~2 weights. The model is trained on speech QA data with a multi-task objective: speech-to-speech QA (primary), speech-to-text QA (auxiliary, weight 5), and text-to-speech QA (auxiliary, weight 1). The contrastive representation alignment loss is applied at all three stages.

\subsection{Speech-to-text}

\begin{figure}[t]
    \centering
    \includegraphics[width=1.0\linewidth]{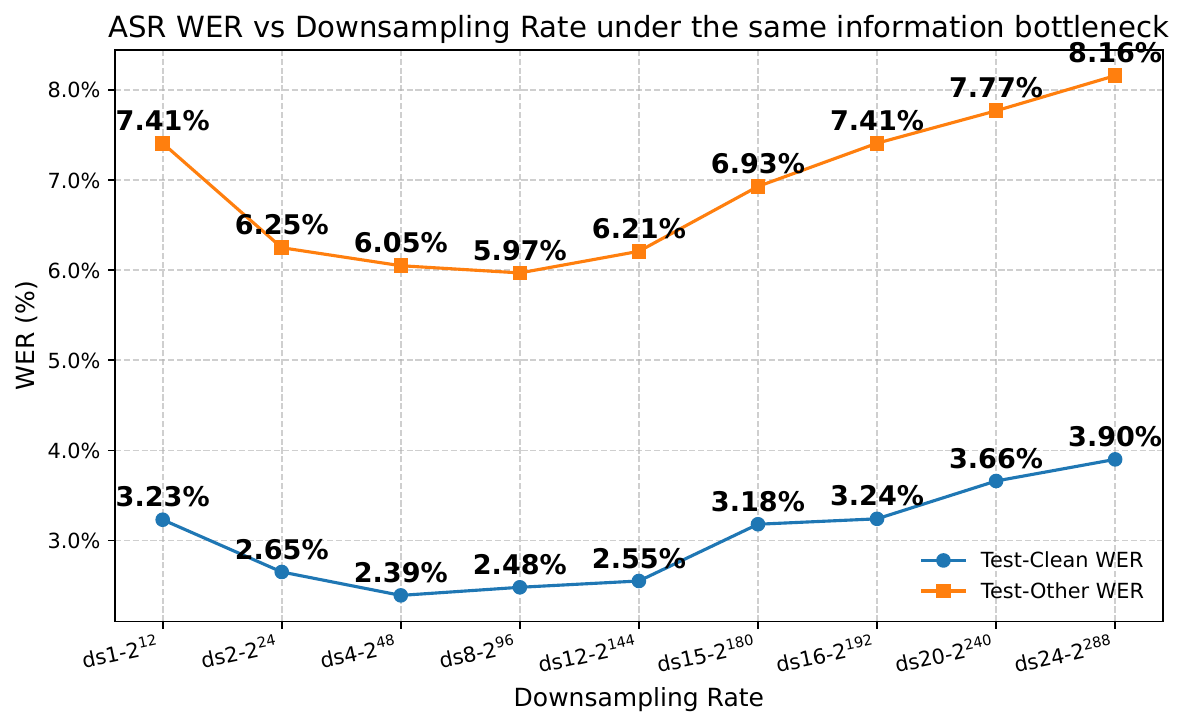}
    \caption{ASR WER at different downsampling rates under fixed information rate.}
    \label{fig:wer_ds}
\end{figure}

All models are trained on LibriSpeech 960h~\cite{panayotov2015librispeech} for 5 epochs (lr $3{\times}10^{-4}$, AdamW optimizer with linear warmup over the first 3\% of steps). The speech encoder is Whisper-Large-v3, whose output features at 50\,Hz are downsampled by a strided convolutional layer and quantized with FSQ. The quantized features are projected into the LLM embedding space via a trained linear layer. We use LibriSpeech-PC~\cite{meister2023librispeech} transcripts preserving case and punctuation, rather than normalized text, as our experiments indicate that text normalization degrades the transferability of text knowledge in downstream speech-to-speech QA---it removes capitalization and punctuation patterns that the text LLM has learned to model. WER is computed on test-clean and test-other using the Whisper normalizer.

Our experiments start from 50\,Hz with a per-frame codebook of $2^{12}{=}4096$, corresponding to $12 {\times} 50 {=} 600$ bits/s. We fix this bitrate across all downsampling factors: at $2{\times}$ (25\,Hz) we use $2^{24}$ (24 bits/frame); at $24{\times}$ (2.08\,Hz) each frame carries 288 bits ($2^{288}$ implicit codebook). The number of FSQ groups $n$ is adjusted to keep each group's prediction at $2^{12}$ categories.

Results (Fig.~\ref{fig:wer_ds}) show ASR WER remains in a narrow range (test-other: 5.97--8.16, test-clean: 2.39--3.90), validating that our factorized FSQ approach resolves the information bottleneck at low frame rates. Compared to the fixed-codebook baselines (Figs.~\ref{fig:singlevqwer_4k}--\ref{fig:singlevqwer_256k}), where WER degrades catastrophically below 6.25\,Hz, this confirms the effectiveness of our scalable codebook.

We observe a U-shaped pattern: when the sequence is excessively long (high frame rate), performance deteriorates as the frozen LLM struggles with redundant temporal resolution---the self-attention mechanism must process many tokens without proportional increase in semantic content, diluting per-position semantic density. Conversely, when the sequence becomes shorter than text (low frame rate), WER increases due to lossy compression at high per-frame information density. The most favorable region emerges at intermediate frame rates (12.5, 6.25, and 4.17\,Hz). This U-shaped trend is consistent across both test sets, suggesting that, under a fixed information rate, there is an optimal range of sequence lengths for a frozen text LLM to understand speech tokens.

\textbf{Tokenizer reconstruction.} Beyond ASR, we evaluate speech reconstruction from tokens on SeedTTS test-en using a token-to-waveform reconstruction model (\textsc{token2wav}). Following prior work~\cite{du2024cosyvoice}, we assess intelligibility (WER), speaker similarity (SIM), and speech quality (UTMOS). SIM is computed as the cosine similarity between WavLM-TDNN embeddings of the reference and reconstructed speech~\cite{chen2024wavlm}. WER is measured using Whisper-large-v3~\cite{radford2023whisper}. Speech quality is evaluated using the official UTMOS checkpoint~\cite{saeki2022utmos}.
For the CosyVoice2 tokenizer baseline, \textsc{token2wav} maps discrete tokens to continuous conditioning sequences via embedding lookup, and upsamples them to 50\,Hz before feeding them as conditional inputs to the flow-matching based acoustic model. For our FSQ tokenizer, we instead deterministically dequantize tokens back to continuous quantized features via the FSQ inverse mapping, and similarly upsample to 50\,Hz for flow-matching conditioning. We train \textsc{token2wav} on Emilia-en using the same architecture and data structure as the CosyVoice baseline.

\begin{table}[t]
\centering
\small
\caption{Tokenizer reconstruction on SeedTTS test-en. The CosyVoice2 tokenizer baseline follows CosyVoice-style supervised semantic tokens~\cite{du2024cosyvoice}.}
\label{tab:tok_recon}
\begin{tabular}{lccc}
\toprule
Tokenizer & WER $\downarrow$ & SIM $\uparrow$ & UTMOS $\uparrow$ \\
\midrule
CosyVoice2 (25\,Hz) & 4.10 & 0.68 & 3.65 \\
Ours (25\,Hz) & 3.04 & 0.67 & 3.71 \\
Ours (4.17\,Hz) & 3.37 & 0.65 & 3.79 \\
\bottomrule
\end{tabular}
\end{table}
Overall, our tokenizer remains readily reconstructable: at 25\,Hz it improves WER and UTMOS over CosyVoice2, and even at 4.17\,Hz it preserves strong speaker similarity and speech quality, indicating that lowering the frame rate does not create a reconstruction bottleneck.
At 25\,Hz, the lower WER of our tokenizer is expected: compared to CosyVoice2's 6561 codebook of semantic token, our grouped FSQ representation can allocate substantially more discrete capacity per frame (e.g., implicit codebook size $2^{48}$), yielding more precise content tokens for reconstruction.

\FloatBarrier
\subsection{Text-to-speech}

\begin{figure}[t]
    \centering
    \includegraphics[width=1.0\linewidth]{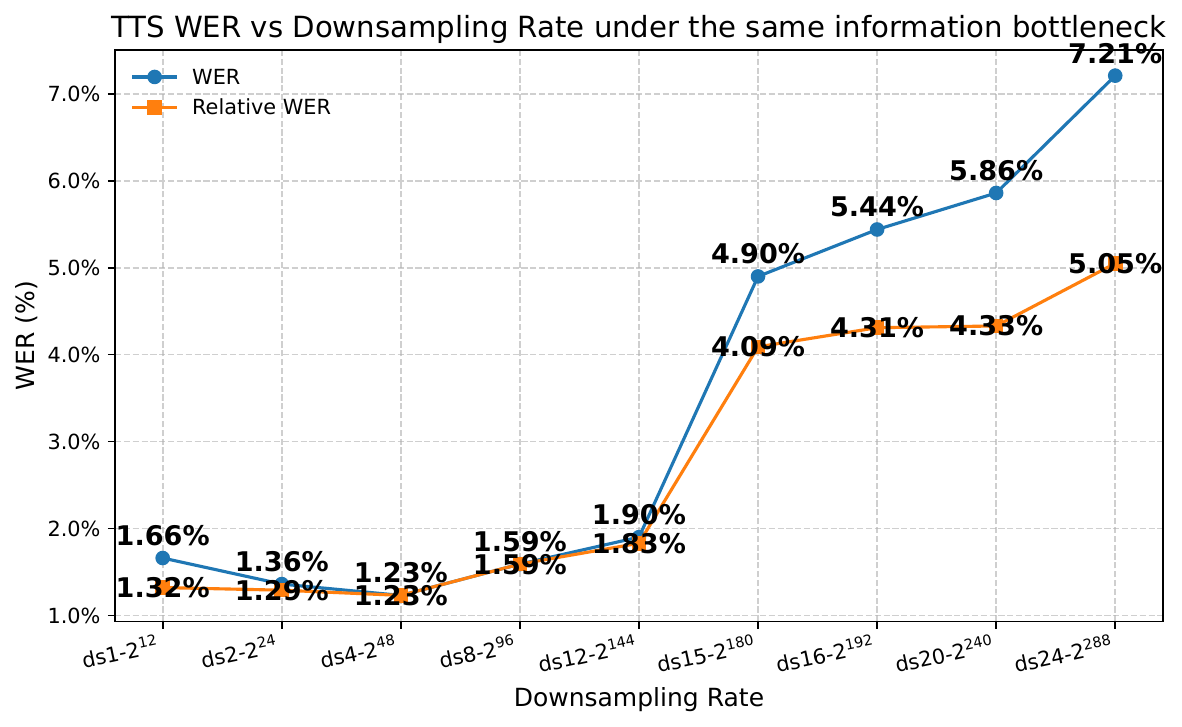}
    \caption{TTS WER at different downsampling rates under fixed information rate.}
    \label{fig:t2s}
\end{figure}

We next train the mapping from text tokens to speech tokens. We reuse the speech tokens from the ASR stage (freezing the downsampling module and quantizer) and train only the projector and audio LM head with the same hyperparameters. Since the speech token sequences are reused from ASR, their text transcriptions are also recognized by the previous stage's ASR model. To further normalize differences in ASR quality, we additionally compute a \emph{relative WER}, where the best-performing configuration is scaled to 1 and others are adjusted proportionally.

\begin{table}[t]
\centering
\small
\caption{Ablation of the audio prediction head for TTS at 4.17\,Hz (ds12), measured by WER. The linear head removes the NAR transformer (inter-group self-attention).}
\label{tab:tts_head_ablation}
\begin{tabular}{lcc}
\toprule
Head & test-clean $\downarrow$ & test-other $\downarrow$ \\
\midrule
Linear (w/o NAR) & 10.17 & 12.73 \\
NAR (ours) & 1.83 & 1.90 \\
\bottomrule
\end{tabular}
\end{table}
As shown in Table~\ref{tab:tts_head_ablation}, removing the NAR transformer and using a linear head increases WER from 1.83/1.90 to 10.17/12.73 at 4.17\,Hz. This confirms that explicitly modeling inter-group dependencies in the audio head is crucial for high-density speech-token prediction at low frame rates.

Results (Fig.~\ref{fig:t2s}) show a consistent trend: as downsampling increases (lower frame rate), WER monotonically worsens. More importantly, once the speech sequence length approaches or becomes shorter than the average text length (${\sim}$3.32 tokens/second), a sharp degradation occurs. This suggests that when speech tokens become much shorter than text tokens, the information density per token is too high for the frozen text LLM to accurately predict them.

Notably, the TTS trend differs from the ASR trend: while ASR shows a U-shaped pattern (degradation at both extremes), TTS degrades monotonically with increasing downsampling. This asymmetry arises because understanding and generation impose different demands on the frozen LLM. For understanding (ASR), the LLM can aggregate semantic content from redundant tokens; for generation (TTS), the LLM must produce the exact codebook indices at each timestep, which becomes increasingly difficult as each token carries more information. The TTS degradation at low frame rates effectively sets a lower bound on viable frame rates for a full dialogue system.  

\subsection{Speech-to-speech QA}

\begin{figure}[t]
    \centering
    \includegraphics[width=1.0\linewidth]{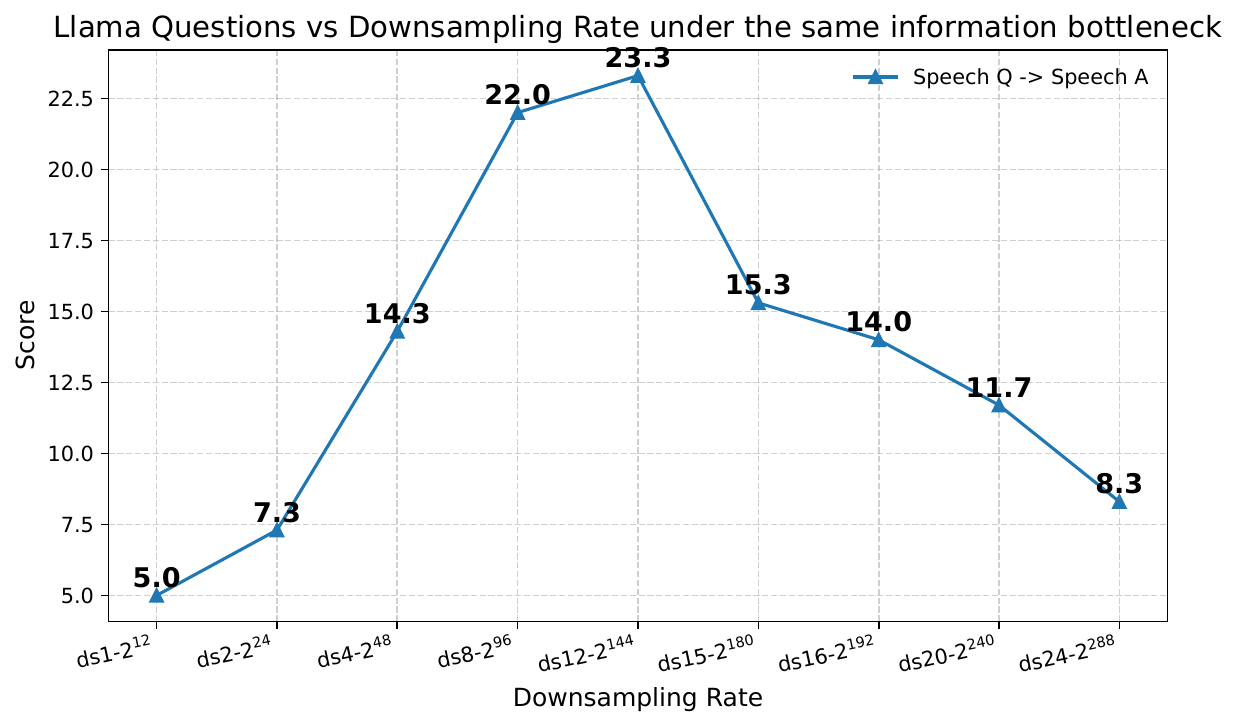}
    \caption{Speech QA results across different downsampling rates under fixed information rate.}
    \label{fig:qa_results}
\end{figure}

In this stage, we focus on speech QA. The speech tokens are reused from the ASR stage, and the model architecture is the same as in the TTS stage. The projector and audio head are initialized from the previous stage. For training, we adopt the InstructS2S-200k dataset~\cite{fang2025llamaomni}, which contains 1,500 hours of speech QA data. During training, we include not only speech-to-speech QA pairs but also the corresponding speech-to-text QA and text-to-speech QA as auxiliary tasks. The loss weight for speech-to-text QA is set to 5, while all other tasks are weighted as 1. We train for 3 epochs with a learning rate of $10^{-4}$.

We evaluate QA on questions from Web Questions, Llama Questions, and TriviaQA, following prior work~\cite{defossez2024moshi,zeng2025scaling}. For each item, we decode the model's speech-token response into a text response using the Stage~1 ASR model, and provide GPT-4o with the ground-truth answer to judge whether the response is correct.  

As shown in Fig.~\ref{fig:qa_results}, the QA scores first increase and then decrease, with clear peaks at 4.17\,Hz and 6.25\,Hz. At 50\,Hz, despite good ASR and TTS performance individually, speech QA is poor because the speech sequence is approximately 15$\times$ longer than text, making cross-modal reasoning difficult for the frozen LLM. At very low frame rates ($<$3.32\,Hz), the model cannot reliably generate such high-density tokens.

The best performance at 4.17\,Hz rather than at 3.32\,Hz deserves explanation. The text token rate of 3.32\,Hz is an \emph{average}: individual utterances vary substantially depending on speaking rate and vocabulary. Operating at exactly 3.32\,Hz means that shorter-than-average utterances produce speech sequences shorter than their text counterparts, which degrades performance sharply as shown by the TTS results. By operating slightly above the mean, the system provides a buffer that keeps even faster-speaking utterances within the LLM's tolerance range. Settings at 4.17\,Hz and 6.25\,Hz both cover this variance effectively.

The inverted-U shape across all three tasks paints a coherent picture: for understanding (ASR), moderate downsampling compresses redundancy without losing content; for generation (TTS), the frozen LLM struggles with high-density tokens at very low rates; for the combined task (QA), the optimum occurs where both sides perform well, around 4--6\,Hz. The ranking of frame rates is consistent across all three QA benchmarks despite different knowledge domains, strengthening the claim that this finding generalizes.

\subsection{Analysis: why slightly above the text rate?}

The finding that 4.17\,Hz outperforms 3.32\,Hz (which exactly matches the average text rate) warrants further analysis. We conduct an auxiliary experiment to probe the frozen LLM's tolerance to input length variation: we stretch Qwen3-4B's text embeddings by linear interpolation at different scaling factors and evaluate on MMLU~\cite{hendrycks2021measuringmassivemultitasklanguage}. While speech-text alignment is not strictly linear, it is monotonic, making this a reasonable approximation.

\begin{figure}[t]
    \centering
    \includegraphics[width=1.0\linewidth]{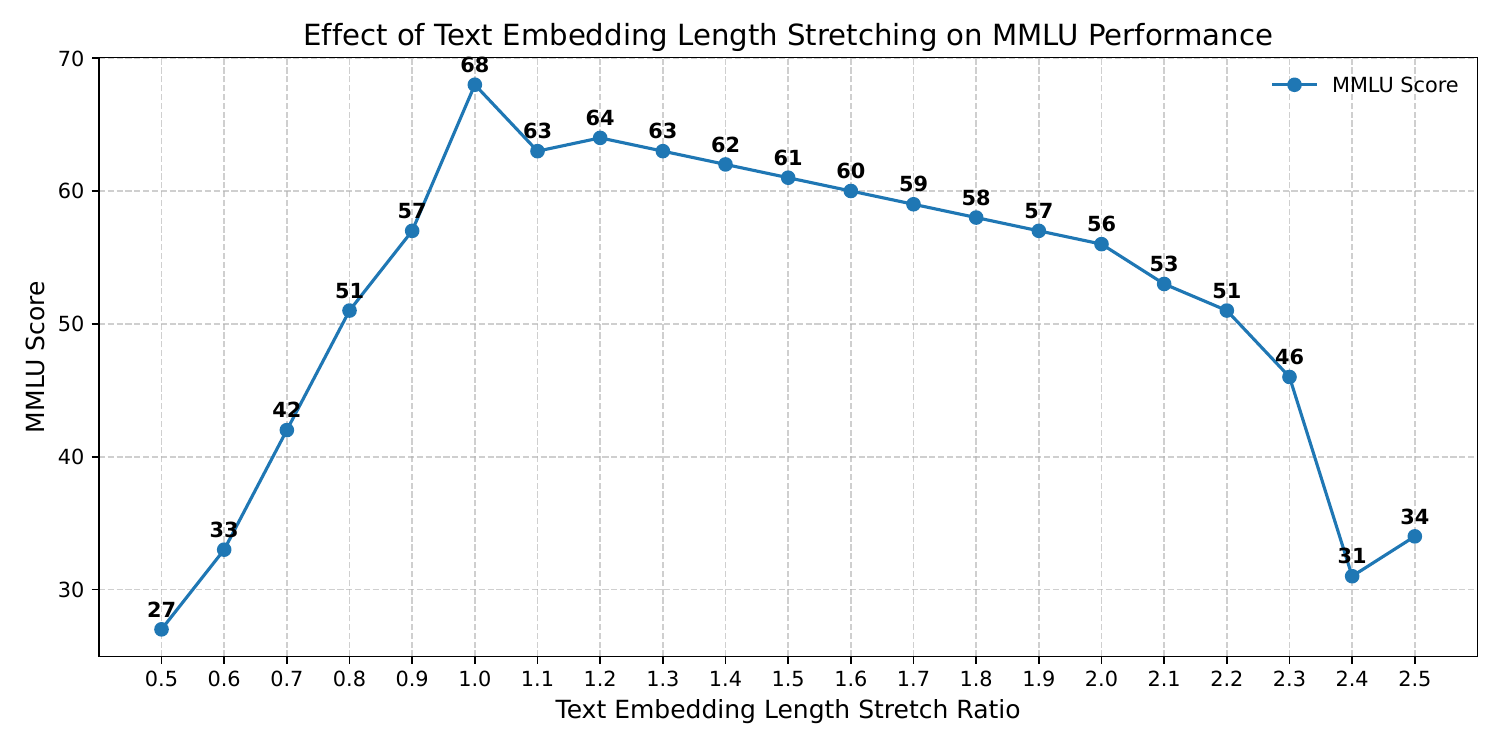}
    \caption{Effect of text embedding length stretching on MMLU performance for the frozen Qwen3-4B backbone.}
    \label{fig:mmlu_stretch}
\end{figure}

As shown in Fig.~\ref{fig:mmlu_stretch}, MMLU accuracy remains above 50\% for scaling factors between 0.8$\times$ and 2.2$\times$, but drops sharply below 0.8$\times$.

This explains the QA results: the text token rate of 3.32\,Hz is an average, masking substantial per-utterance variance. At the 3.13\,Hz (ds16) setting, a significant fraction of utterances produce speech sequences shorter than 0.8$\times$ their text length, falling outside the LLM's tolerance. In contrast, 4.17\,Hz and 6.25\,Hz keep almost all utterances within the 0.8--2.2$\times$ tolerance window, explaining their superior performance. This analysis provides a principled criterion for choosing the frame rate: target a rate that keeps the speech-to-text length ratio within the LLM's verified tolerance range across the entire utterance distribution, rather than simply matching the average.

\subsection{Ablation: speech encoder}

We ablate the frozen speech feature extractor while keeping the rest of the setup fixed (4.17\,Hz and middle-layer representation alignment). As shown in Table~\ref{tab:encoder_ablation}, multiple speech encoders work in this framework, while stronger encoders yield higher QA performance. Whisper's advantage likely stems from its ASR-supervised features being inherently closer to text semantics, which facilitates cross-modal alignment; nonetheless, the framework remains effective with self-supervised encoders (HuBERT, WavLM).

\begin{table}[t]
\centering
\small
\caption{Effect of speech encoder on Llama Questions (speech QA score).}
\label{tab:encoder_ablation}
\begin{tabular}{lc}
\toprule
Speech encoder & Llama Q. \\
\midrule
Whisper-Large-v3 & 30.7 \\
HuBERT-Large & 25.7 \\
WavLM-Large & 27.3 \\
\bottomrule
\end{tabular}
\end{table}

\subsection{Representation alignment experiments}

We incorporate the contrastive alignment loss at all stages of training (ASR, TTS, and speech QA). For each speech clip, we extract the hidden representation of its corresponding transcription through the frozen text LLM. We select a probe layer $\ell$ and average the hidden states over the temporal axis to obtain vectors $\widehat{h}_s, \widehat{h}_t \in \mathbb{R}^d$ (L2-normalized). We then optimize the InfoNCE objective with temperature $\tau{=}0.07$.

We investigate which LLM layer provides the most effective alignment signal by sweeping over four options: the text input embedding (\textsc{Emb}), an early layer ($\lfloor L/4\rfloor$), a middle layer ($\lfloor L/2\rfloor$), and a late layer ($\lfloor 3L/4\rfloor$), using the 4.17\,Hz frame rate setting.

\begin{table}[t]
\caption{Effect of alignment layer on Llama Questions (speech QA score). Middle layer ($L/2$) yields the best result.}
\label{tab:ssqa_layer}
\centering
\small
\begin{tabular}{lccccc}
\toprule
& \textbf{None} & \textbf{Emb} & $\mathbf{L/4}$ & $\mathbf{L/2}$ & $\mathbf{3L/4}$ \\
\midrule
Llama Q. & 23.3 & 21.7 & 25.0 & 30.7 & 27.7 \\
\bottomrule
\end{tabular}
\end{table}

As shown in Table~\ref{tab:ssqa_layer}, alignment at the middle layer ($L/2$) provides the largest improvement (+7.4 points over no alignment), consistent with findings that intermediate LLM layers encode the most transferable semantic representations~\cite{lyu2024dmcodec,lei2023ckdst}. Embedding-level alignment slightly degrades performance ($-$1.6 points), suggesting the modality gap primarily lies in deeper representations rather than the input embedding space---the input projector already maps speech tokens into the embedding space, so the more meaningful gap exists in representations after several attention layers. Late-layer alignment ($3L/4$) improves over the baseline (+4.4 points) but is less effective than mid-layer, possibly because later layers are more specialized for text-specific generation patterns. The early layer ($L/4$) provides a moderate improvement (+1.7 points).  

\subsection{Comparison with other methods}

\begin{table}[t]
\centering
\small
\caption{Speech QA results (S2S only) on Web Questions, Llama Questions, and Trivia QA. We compare pure speech-token response systems.}
\label{tab:comparison}
\resizebox{\linewidth}{!}{%
\begin{tabular}{llcccc}
\toprule
Model & Train Params & Data & Web Q. & Llama Q. & Trivia QA \\
\midrule
\multicolumn{6}{l}{\textit{Moshi}~\cite{defossez2024moshi}} \\
7B & 7B & 7M hrs & 9.2 & 21.0 & 7.3 \\
\midrule
\multicolumn{6}{l}{\textit{Scaling Interleave}~\cite{zeng2025scaling}} \\
9B & 9B & 600B tok & 15.9 & 50.7 & 26.5 \\
9B & 9B & 200B tok & 13.3 & 44.0 & 18.7 \\
9B & 9B & 100B tok & 9.3 & 37.0 & 11.7 \\
\midrule
\multicolumn{6}{l}{\textit{Ours}} \\
4B & ${\sim}$100M & 2.5k hrs & 7.9 & 30.7 & 11.9 \\
8B & ${\sim}$150M & 2.5k hrs & 12.2 & 39.3 & 17.6 \\
\bottomrule
\end{tabular}%
}
\end{table}

We further train a model based on Qwen3-8B with the same settings as the 4B version: 4.17\,Hz frame rate and middle-layer alignment. The 8B model uses a larger hidden dimension (3584 vs.\ 2560), resulting in ${\sim}$150M trainable parameters for the projector and audio head.

As shown in Table~\ref{tab:comparison}, our 4B model---trained on only 2.5k hours with ${\sim}$100M trainable parameters---outperforms Moshi~\cite{defossez2024moshi} (7B backbone, 7M hours, full-parameter training) across all three QA benchmarks. Our 8B model achieves results comparable to systems requiring 100B--200B interleaved tokens and full 9B parameter training~\cite{zeng2025scaling}, despite using only 2.5k hours and updating ${\sim}$150M parameters. The consistent scaling from 4B to 8B (+4.3 on Web Q., +8.6 on Llama Q., +5.7 on Trivia QA) suggests our alignment approach effectively leverages stronger backbones, and allows immediate upgrade by simply swapping the frozen backbone.
We emphasize that Table~\ref{tab:comparison} reports the \emph{pure} speech-to-speech setting (S2S only), where the model must generate speech tokens without producing an intermediate text response. We therefore do not directly compare against thinker--talker style systems that first generate a text answer and then condition speech synthesis on that text, since such text guidance changes the task definition and can mask the difficulty of directly speech-token generation.

\subsection{Discussion}

Our experiments, conducted under a frozen text LLM backbone with a fixed information budget, isolate how speech-token temporal structure and representation alignment affect cross-modal transfer.

\textbf{Frame rate as a first-class design variable.} Prior work has largely treated the speech tokenizer's frame rate as a fixed hyperparameter. Our sweep from 50\,Hz to 2.08\,Hz shows that frame rate is one of the most impactful choices, with speech QA scores differing by up to $4{\times}$ across frame rates under the same information budget. The optimal rate (${\sim}$4.17\,Hz) closely tracks the text token rate.

\textbf{Information rate vs.\ frame rate.} Controlling the information throughput is necessary but not sufficient. Even with matched bits/second, performance varies dramatically with frame rate, confirming that the \emph{temporal granularity} of the token sequence itself matters for frozen-LLM compatibility.

\textbf{Synergy between length and representation alignment.} Length alignment alone at 4.17\,Hz yields 23.3 on Llama Questions; adding mid-layer representation alignment improves it to 30.7 ($+$32\% relative). Since the backbone is frozen, this gain reflects improved representation compatibility rather than LLM adaptation. The two dimensions are complementary: length alignment ensures structural compatibility, while representation alignment closes the semantic gap.

\textbf{Practical guidelines for speech token design.} Our findings suggest: (1)~target moderate frame rates slightly above the \emph{mean} text token rate (4.17--6.25\,Hz in our sweep) to accommodate utterance-level variability; (2)~mitigate temporal-granularity mismatch by compressing redundant frames: long redundant streams tend to spread attention mass and weaken decision signals, while overly aggressive compression increases bits/frame and risks losing semantic content; and (3)~align at intermediate LLM layers via contrastive learning. We expect these trends to transfer to other backbones, though the exact optimum may shift with model and data.

\section{Limitations}

Our study has several limitations.
First, the comparison in Table~\ref{tab:comparison} involves different LLM backbones across methods; since neither the training data nor the training code of the baselines is publicly available, reproducing them on the same backbone is not feasible. The comparison therefore primarily reflects data efficiency rather than a controlled method-to-method evaluation.
Second, the frozen LLM may impose a performance ceiling; fine-tuning the backbone could narrow the gap to fully-trained systems.
Third, our experiments use only English read speech (LibriSpeech) and InstructS2S-200k; generalization to noisy, conversational, or multilingual speech remains to be validated.
Fourth, all findings are based on the Qwen3 family (4B and 8B); whether the optimal frame rate and alignment layer transfer to other LLM architectures requires further verification.
Finally, the reliance on Whisper-Large-v3 as the speech encoder and the lack of acoustic modeling are important directions for future work.

\section{Conclusion}

We presented a controlled study of which speech representation best matches text-native reasoning in a frozen LLM, investigating frame rate and representation alignment as two key design variables under a fixed information throughput. For frame rate, the best regime for speech QA occurs at 4.17--6.25\,Hz, enabled by our scalable FSQ with factorized group prediction and a NAR transformer head that scales per-frame capacity to ${\sim}$300 bits while adding only ${\sim}$100--150M trainable parameters. For representation alignment, InfoNCE at the middle LLM layer yields the strongest gain (+7.4 points), and the two alignment dimensions are complementary (+32\% relative improvement when combined). The resulting frozen-LLM system achieves competitive speech QA with 2.5k hours of data, outperforming systems trained on orders of magnitude more data. These findings provide practical guidelines for speech token design: use factorized codebooks, model inter-group dependencies in the audio head, and align at intermediate layers. We further show that the framework generalizes across speech encoders (Whisper, HuBERT, WavLM) and that our tokenizer maintains strong reconstruction quality even at 4.17\,Hz.

\nocite{wang2024freezeomni,tang2024salmonn,hu2024wavllm,nachmani2024spectron,rubenstein2023audiopalm,zhang2024speechlm,chen2024wavlm,baevski2020wav2vec2,radford2023whisper,borsos2023audiolm,kharitonov2023speak,maiti2024voxtlm,wu2023decoder,hassid2023twist,lyth2024natural,oord2018representation}

\bibliographystyle{IEEEtran}
\bibliography{refs}

@article{zeng2024glm,
  title={{GLM-4-Voice}: Towards intelligent and human-like end-to-end spoken chatbot},
  author={Zeng, Aohan and Du, Zhengxiao and Liu, Mingdao and Wang, Kedong and Jiang, Shengmin and Zhao, Lei and Dong, Yuxiao and Tang, Jie},
  journal={arXiv preprint arXiv:2412.02612},
  year={2024}
}

@article{zhang2023speechgpt,
  title={{SpeechGPT}: Empowering large language models with intrinsic cross-modal conversational abilities},
  author={Zhang, Dong and Li, Shimin and Zhang, Xin and Zhan, Jun and Wang, Pengyu and Zhou, Yaqian and Qiu, Xipeng},
  journal={arXiv preprint arXiv:2305.11000},
  year={2023}
}

@article{chen2024slamomni,
  title={{SLAM-Omni}: Timbre-controllable voice interaction system with single-stage training},
  author={Chen, Wenxi and Ma, Ziyang and Yan, Ruiqi and others},
  journal={arXiv preprint arXiv:2412.15649},
  year={2024}
}

@article{kumar2023dac,
  title={High-fidelity audio compression with improved {RVQGAN}},
  author={Kumar, Rithesh and Seetharaman, Prem and Luebs, Alejandro and Kumar, Ishaan and Kumar, Kundan},
  journal={Advances in Neural Information Processing Systems},
  volume={36},
  pages={27980--27993},
  year={2023}
}

@article{defossez2022encodec,
  title={High fidelity neural audio compression},
  author={D{\'e}fossez, Alexandre and Copet, Jade and Synnaeve, Gabriel and Adi, Yossi},
  journal={arXiv preprint arXiv:2210.13438},
  year={2022}
}

@article{hsu2021hubert,
  title={{HuBERT}: Self-supervised speech representation learning by masked prediction of hidden units},
  author={Hsu, Wei-Ning and Bolte, Benjamin and Tsai, Yao-Hung Hubert and Lakhotia, Kushal and Salakhutdinov, Ruslan and Mohamed, Abdelrahman},
  journal={IEEE/ACM Trans.\ Audio, Speech, Lang.\ Process.},
  volume={29},
  pages={3451--3460},
  year={2021}
}

@inproceedings{lee2022rvq,
  title={Autoregressive image generation using residual quantization},
  author={Lee, Doyup and Kim, Chiheon and Kim, Saehoon and Cho, Minsu and Han, Wook-Shin},
  booktitle={Proc. IEEE/CVF CVPR},
  pages={11523--11532},
  year={2022}
}

@article{xie2024miniomni,
  title={{Mini-Omni}: Language models can hear, talk while thinking in streaming},
  author={Xie, Zhifei and Wu, Changqiao},
  journal={arXiv preprint arXiv:2408.16725},
  year={2024}
}

@article{nguyen2025spirit,
  title={{Spirit-LM}: Interleaved spoken and written language model},
  author={Nguyen, Tu Anh and Muller, Benjamin and Yu, Bokai and others},
  journal={Trans.\ Assoc.\ Comput.\ Linguistics},
  volume={13},
  pages={30--52},
  year={2025}
}

@article{hassid2023textually,
  title={Textually pretrained speech language models},
  author={Hassid, Michael and Remez, Tal and Nguyen, Tu Anh and Gat, Itai and Conneau, Alexis and Kreuk, Felix and Copet, Jade and Defossez, Alexandre and Synnaeve, Gabriel and Dupoux, Emmanuel and others},
  journal={Advances in Neural Information Processing Systems},
  volume={36},
  pages={63483--63501},
  year={2023}
}

@article{lakhotia2021generative,
  title={On generative spoken language modeling from raw audio},
  author={Lakhotia, Kushal and Kharitonov, Eugene and Hsu, Wei-Ning and others},
  journal={Trans.\ Assoc.\ Comput.\ Linguistics},
  volume={9},
  pages={1336--1354},
  year={2021}
}

@inproceedings{panayotov2015librispeech,
  title={{LibriSpeech}: An {ASR} corpus based on public domain audio books},
  author={Panayotov, Vassil and Chen, Guoguo and Povey, Daniel and Khudanpur, Sanjeev},
  booktitle={Proc. IEEE ICASSP},
  pages={5206--5210},
  year={2015}
}

@inproceedings{meister2023librispeech,
  title={{LibriSpeech-PC}: Benchmark for evaluation of punctuation and capitalization capabilities of end-to-end {ASR} models},
  author={Meister, Aleksandr and Novikov, Matvei and Karpov, Nikolay and Bakhturina, Evelina and Lavrukhin, Vitaly and Ginsburg, Boris},
  booktitle={Proc. IEEE ASRU},
  pages={1--7},
  year={2023}
}

@article{yang2025qwen3,
  title={Qwen3 technical report},
  author={Yang, An and Li, Anfeng and Yang, Baosong and others},
  journal={arXiv preprint arXiv:2505.09388},
  year={2025}
}

@article{du2024cosyvoice,
  title={{CosyVoice}: A scalable multilingual zero-shot text-to-speech synthesizer based on supervised semantic tokens},
  author={Du, Zhihao and Chen, Qian and Zhang, Shiliang and others},
  journal={arXiv preprint arXiv:2407.05407},
  year={2024}
}

@article{fang2024llama,
  title={{LLaMA-Omni}: Seamless speech interaction with large language models},
  author={Fang, Qingkai and Guo, Shoutao and Zhou, Yan and Ma, Zhengrui and Zhang, Shaolei and Feng, Yang},
  journal={arXiv preprint arXiv:2409.06666},
  year={2024}
}

@article{defossez2024moshi,
  title={Moshi: A speech-text foundation model for real-time dialogue},
  author={D{\'e}fossez, Alexandre and Mazar{\'e}, Laurent and Orsini, Manu and others},
  journal={arXiv preprint arXiv:2410.00037},
  year={2024}
}

@inproceedings{zeng2025scaling,
  title={Scaling speech-text pre-training with synthetic interleaved data},
  author={Zeng, Aohan and Du, Zhengxiao and Liu, Mingdao and Zhang, Lei and Jiang, Shengmin and Dong, Yuxiao and Tang, Jie},
  booktitle={Proc. ICLR},
  year={2025}
}

@inproceedings{fang2025llamaomni,
  title={{LLaMA-Omni}: Seamless speech interaction with large language models},
  author={Fang, Qingkai and Guo, Shoutao and Zhou, Yan and Ma, Zhengrui and Zhang, Shaolei and Feng, Yang},
  booktitle={Proc. ICLR},
  year={2025}
}

@article{xu2025qwen2,
  title={{Qwen2.5-Omni} technical report},
  author={Xu, Jin and Guo, Zhifang and He, Jinzheng and others},
  journal={arXiv preprint arXiv:2503.20215},
  year={2025}
}

@article{ding2025kimi,
  title={{Kimi-Audio} technical report},
  author={Ding, Ding and Ju, Zeqian and Leng, Yichong and others},
  journal={arXiv preprint arXiv:2504.18425},
  year={2025}
}

@article{huang2025step,
  title={{Step-Audio}: Unified understanding and generation in intelligent speech interaction},
  author={Huang, Ailin and Wu, Boyong and Wang, Bruce and others},
  journal={arXiv preprint arXiv:2502.11946},
  year={2025}
}

@misc{mentzer2023fsq,
  title={Finite scalar quantization: {VQ-VAE} made simple},
  author={Mentzer, Fabian and Minnen, David and Agustsson, Eirikur and Tschannen, Michael},
  year={2023},
  eprint={2309.15505},
  archivePrefix={arXiv}
}

@article{wang2024blsp,
  title={Blsp: Bootstrapping language-speech pre-training via behavior alignment of continuation writing},
  author={Wang, Chen and Liao, Minpeng and Huang, Zhongqiang and Lu, Jinliang and Wu, Junhong and Liu, Yuchen and Zong, Chengqing and Zhang, Jiajun},
  journal={arXiv preprint arXiv:2309.00916},
  year={2023}
}

@inproceedings{fathullah2024audiochatllama,
  title={Audiochatllama: Towards general-purpose speech abilities for llms},
  author={Fathullah, Yassir and Wu, Chunyang and Lakomkin, Egor and Li, Ke and Jia, Junteng and Shangguan, Yuan and Mahadeokar, Jay and Kalinli, Ozlem and Fuegen, Christian and Seltzer, Mike},
  booktitle={Proceedings of the 2024 Conference of the North American Chapter of the Association for Computational Linguistics: Human Language Technologies (Volume 1: Long Papers)},
  pages={5522--5532},
  year={2024}
}

@inproceedings{held2024diva,
  title={Distilling an end-to-end voice assistant without instruction training data},
  author={Held, William and Zhang, Yanzhe and Li, Minzhi and Shi, Weiyan and Ryan, Michael J and Yang, Diyi},
  booktitle={Proceedings of the 63rd Annual Meeting of the Association for Computational Linguistics (Volume 1: Long Papers)},
  pages={7876--7891},
  year={2025}
}

@article{lyu2024dmcodec,
  title={Dm-codec: Distilling multimodal representations for speech tokenization},
  author={Ahasan, Md Mubtasim and Fahim, Md and Mohiuddin, Tasnim and Rahman, AKM and Chadha, Aman and Iqbal, Tariq and Amin, M Ashraful and Islam, Md Mofijul and Ali, Amin Ahsan},
  journal={arXiv preprint arXiv:2410.15017},
  year={2024}
}

@inproceedings{vandenoord2017vqvae,
  title={Neural discrete representation learning},
  author={van den Oord, Aaron and Vinyals, Oriol and Kavukcuoglu, Koray},
  booktitle={Advances in Neural Information Processing Systems},
  volume={30},
  year={2017}
}

@article{ju2024naturalspeech3,
  title={{NaturalSpeech 3}: Zero-shot speech synthesis with factorized codec and diffusion models},
  author={Ju, Zeqian and Wang, Yuancheng and Shen, Kai and others},
  journal={arXiv preprint arXiv:2403.03100},
  year={2024}
}

@article{zhang2024speechtokenizer,
  title={{SpeechTokenizer}: Unified speech tokenizer for speech large language models},
  author={Zhang, Xin and Zhang, Dong and Li, Shimin and others},
  journal={arXiv preprint arXiv:2308.16692},
  year={2023}
}

@inproceedings{lei2023ckdst,
  title={{CKDST}: Comprehensively and effectively distill knowledge from machine translation to end-to-end speech translation},
  author={Lei, Yikun and others},
  booktitle={Findings of ACL},
  year={2023}
}

@article{wang2024freezeomni,
  title={{Freeze-Omni}: A smart and low latency speech-to-speech dialogue model with frozen {LLM}},
  author={Wang, Xiong and Zhu, Yangze and Yu, Haoxiang and others},
  journal={arXiv preprint arXiv:2411.00774},
  year={2024}
}

@inproceedings{tang2024salmonn,
  title={{SALMONN}: Towards generic hearing abilities for large language models},
  author={Tang, Changli and Yu, Wenyi and Sun, Guangzhi and others},
  booktitle={Proc. ICLR},
  year={2024}
}

@article{hu2024wavllm,
  title={{WavLLM}: Towards robust and adaptive speech large language model},
  author={Hu, Shujie and Zhou, Long and Liu, Shujie and others},
  journal={arXiv preprint arXiv:2404.00656},
  year={2024}
}

@article{nachmani2024spectron,
  title={Spoken question answering and speech continuation using spectrogram-powered {LLM}},
  author={Nachmani, Eliya and Levkovitch, Alon and Hirsch, Roy and others},
  journal={arXiv preprint arXiv:2305.15255},
  year={2023}
}

@article{rubenstein2023audiopalm,
  title={{AudioPaLM}: A large language model that can speak and listen},
  author={Rubenstein, Paul K and Asawaroengchai, Chulayuth and Nguyen, Duc Dung and others},
  journal={arXiv preprint arXiv:2306.12925},
  year={2023}
}

@article{zhang2024speechlm,
  title={{SpeechLM}: Enhanced speech pre-training with unpaired textual data},
  author={Zhang, Ziqiang and Chen, Sanyuan and Zhou, Long and others},
  journal={IEEE/ACM Trans.\ Audio, Speech, Lang.\ Process.},
  volume={32},
  pages={1602--1616},
  year={2024}
}

@article{chen2024wavlm,
  title={{WavLM}: Large-scale self-supervised pre-training for full stack speech processing},
  author={Chen, Sanyuan and Wang, Chengyi and Chen, Zhengyang and others},
  journal={IEEE J.\ Sel.\ Topics Signal Process.},
  volume={16},
  number={6},
  pages={1505--1518},
  year={2022}
}

@inproceedings{saeki2022utmos,
  title={{UTMOS}: UTokyo-SaruLab system for VoiceMOS Challenge 2022},
  author={Saeki, Takaaki and Xin, Detai and Nakata, Wataru and Koriyama, Tomoki and Takamichi, Shinnosuke and Saruwatari, Hiroshi},
  booktitle={Proc. Interspeech},
  pages={4521--4525},
  year={2022}
}

@article{baevski2020wav2vec2,
  title={wav2vec 2.0: A framework for self-supervised learning of speech representations},
  author={Baevski, Alexei and Zhou, Yuhao and Mohamed, Abdelrahman and Auli, Michael},
  journal={Advances in Neural Information Processing Systems},
  volume={33},
  pages={12449--12460},
  year={2020}
}

@inproceedings{radford2023whisper,
  title={Robust speech recognition via large-scale weak supervision},
  author={Radford, Alec and Kim, Jong Wook and Xu, Tao and Brockman, Greg and McLeavey, Christine and Sutskever, Ilya},
  booktitle={Proc. ICML},
  pages={28492--28518},
  year={2023}
}

@article{borsos2023audiolm,
  title={{AudioLM}: A language modeling approach to audio generation},
  author={Borsos, Zal{\'a}n and Marinier, Rapha{\"e}l and Vincent, Damien and others},
  journal={IEEE/ACM Trans.\ Audio, Speech, Lang.\ Process.},
  volume={31},
  pages={2523--2533},
  year={2023}
}

@article{kharitonov2023speak,
  title={Speak, read and prompt: High-fidelity text-to-speech with minimal supervision},
  author={Kharitonov, Eugene and Vincent, Damien and Borsos, Zal{\'a}n and others},
  journal={Trans.\ Assoc.\ Comput.\ Linguistics},
  volume={11},
  pages={1703--1718},
  year={2023}
}

@inproceedings{maiti2024voxtlm,
  title={Voxtlm: Unified decoder-only models for consolidating speech recognition, synthesis and speech, text continuation tasks},
  author={Maiti, Soumi and Peng, Yifan and Choi, Shukjae and Jung, Jee-weon and Chang, Xuankai and Watanabe, Shinji},
  booktitle={ICASSP 2024-2024 IEEE International Conference on Acoustics, Speech and Signal Processing (ICASSP)},
  pages={13326--13330},
  year={2024},
  organization={IEEE}
}

@article{wu2023decoder,
  title={Decoder-only architecture for speech recognition with {CTC} prompts and text-only training},
  author={Wu, Hayato and Yan, Brian and Shi, Yosuke and Watanabe, Shinji},
  journal={arXiv preprint arXiv:2309.08876},
  year={2023}
}

@article{lyth2024natural,
  title={Natural language guidance for controllable {TTS}},
  author={Lyth, Dan and King, Simon},
  journal={arXiv preprint arXiv:2402.01912},
  year={2024}
}

@article{oord2018representation,
  title={Representation learning with contrastive predictive coding},
  author={van den Oord, Aaron and Li, Yazhe and Vinyals, Oriol},
  journal={arXiv preprint arXiv:1807.03748},
  year={2018}
}

@article{hendrycks2021measuringmassivemultitasklanguage,
  title={Measuring Massive Multitask Language Understanding},
  author={Hendrycks, Dan and Burns, Collin and Basart, Steven and Zou, Andy and Mazeika, Mantas and Song, Dawn and Steinhardt, Jacob},
  journal={Proceedings of the International Conference on Learning Representations (ICLR)},
  year={2021}
}

\section{Generative AI Use Disclosure}
A large language model was employed solely for language refinement, including grammar, spelling, clarity, and tone, on text originally crafted by the authors. The LLM did not introduce substantive changes to claims, data interpretation, or conclusions.

\end{document}